\documentclass[twocolumn,10pt]{article}
\usepackage{aas}
\title{
    Meteor statistics I\\
    The distribution of instrumental magnitudes}

\author[1]{Althea V.\ Moorhead}
\affil[1]{NASA Meteoroid Environment Office, Marshall Space Flight Center, Huntsville, AL 35812, USA; \href{althea.moorhead@nasa.gov}{althea.moorhead@nasa.gov}}

\author[2]{Peter G.\ Brown}
\author[2]{Margaret D.\ Campbell-Brown}
\affil[2]{Department of Physics and Astronomy, University of Western Ontario, London, ON N6A 3K7, Canada}

\author[3]{Michael J.\ Mazur}
\affil[3]{Department of Earth Sciences, Western University, 1151 Richmond St. London, Ontario Canada N6A 3K7, Canada}

\author[2,4]{Denis Vida}
\affil[4]{Institute for Earth and Space Exploration, University of Western Ontario, London, ON N6A 5B8, Canada}

\begin{document}

\section{Introduction}
\label{sec:intro}

Meteor magnitudes are commonly assumed to follow an exponential distribution \citep{watson39}; that is, the number of meteors brighter than magnitude $M$ is
\begin{align}
    N_M &\propto r^M = e^{M \ln r} \, .
    \label{eq:nm1}
\end{align}
In practice, the sensitivity of the observational system truncates this distribution by placing an upper limit on $M$.

\subsection{Population index}

The base of the exponent in eq.\,\ref{eq:nm1}, $r$, is called the population index. Its value varies between meteor populations: meteor showers tend to have smaller values of $r$ than the sporadic complex, and $r$ also tends to differ between meteor showers \citep[see][for a review]{moorhead25}.
Meteor magnitudes are often used to estimate meteoroid masses, and so the population index has an equivalent mass index, $s$. Suppose that absolute magnitude is related to mass ($m$) as follows:
\begin{align}
    M &= -2.5 B \log_{10} m + C \, ,
    \label{eq:magmass}
\end{align}
where $B$ and $C$ are constants. By combining eqs.\,\ref{eq:nm1} and \ref{eq:magmass}, we obtain the following expression for the number of meteors with masses greater than $m$:
\begin{align}
    N_m &\propto m^{1 - s} \, , \\
    s &= 1 + 2.5 B \, \log_{10} r \, ,
\end{align}
where $s$ is the differential mass index and ${1-s}$ is the cumulative mass index \citep[see also][]{browne56,shustov22}. Commonly used values of $B$ are 0.9 \citep{jacchia67}, 0.92 \citep{verniani73}, and 1 \citep{campbellbrown16}, which respectively translate to:
\begin{align}
    s &= 1 + 2.25 \log_{10} r \, \text{ for } B = 0.9 \, , \\
    s &= 1 + \hphantom{0}2.3 \log_{10} r \, \text{ for } B = 0.92 \, \text{, and}\\
    s &= 1 + \hphantom{0}2.5 \log_{10} r \, \text{ for } B = 1 \, .
\end{align}
In this paper, we sidestep this ambiguity in $B$ by working in terms of magnitude.

It is necessary to know the population index in order to convert meteor counts to magnitude- or mass-limited fluxes \citep{kaiser60,koschack90a,molau22,vida22}, to compare rates recorded by meteor networks with differing sensitivity \citep{campbellbrown16,blaauw17,ehlert20}, to estimate the total mass contained within a stream \citep{blaauw17}, and to predict visual activity or impact rate using the results of meteoroid stream simulations \citep{moorhead19,egal20}. 

The population index is most commonly measured by fitting a line to the logarithm of the cumulative distribution of meteor magnitude \citep[e.g.,]{hawkins58}. The logarithm of the estimated initial mass \citep[e.g.,]{koten23,barghini24} or radar echo amplitude \citep[e.g.,]{blaauw11,pokorny16,kipreos25} can be used in lieu of magnitude. This can be extended to multiple networks \citep{campbellbrown16,blaauw17} or subnetworks \citep{molau14} by examining the fluxes measured by instruments with differing limiting magnitudes.

\subsection{Detection threshold}

The detection threshold is sometimes measured separately by fitting a distribution to the magnitude data \citep[e.g.]{blaauw11}.
It is much less common to simultaneously measure the detection threshold and population index by fitting the entire magnitude distribution; \cite{betzler15} and \cite{vida20} are two rare examples. \citeauthor{betzler15}\ use an exponentiated generalized Pareto distribution to fit the full magnitude distribution (see sec.\,\ref{sec:egpd}), while \citeauthor{vida20}\ use a gamma distribution (see sec.\,\ref{sec:gamma}). We favor this latter approach, but it requires a model that fits the data well: otherwise, fitting methods may sacrifice accuracy in population index in order to better fit the manner in which observations decrease at dim magnitudes. 

We propose the use of an exponentially modified Gaussian (exGaussian) distribution for modeling meteor magnitudes. In sec.\,\ref{sec:deriv}, we relate the parameters of an exGaussian distribution to population index, detection threshold, and measurement error. In sec.\,\ref{sec:apply}, we show that an exGaussian distribution provides an excellent fit to both the magnitude distribution of faint optical meteors detected by EMCCD cameras at the Canadian Automated Meteor Observatory (CAMO; sec.\,\ref{sec:emccd}) and to the amplitude distribution of radar meteor echoes recorded by the Canadian Meteor Orbit Radar (CMOR; sec.\,\ref{sec:cmor}).

\section{Review of competing models}
\label{sec:dists}

We found at least four different univariate distributions that have been applied to meteor magnitudes in the literature. We review each distribution and its behavior at bright and dim magnitudes. In some cases, we tweak the form of the probability distribution function (PDF) so that the following holds for all distributions:
\begin{align}
   f_M(M) &= \frac{1}{\delta} f_y(y) \, \text{,\, where }\\
   y &= -(M - \mu)/\delta \, .
\end{align}
Here, $M$ is meteor magnitude, $\mu$ is a location parameter, and $\delta$ is a scaling parameter.

\subsection{Gumbel distribution}
\label{sec:gumbel}

\cite{blaauw16} fit a Gumbel distribution to the distribution of stellar magnitudes observed by NASA wide-field meteor cameras; similarly, \cite{ehlert20} fit a Gumbel distribution to the distribution of absolute meteor magnitudes observed by the NASA All Sky Fireball Network. In both cases, the mode of the distribution was taken to be the network's detection or completeness threshold.

\begin{figure*}
    \centering
    \includegraphics[width=\linewidth]{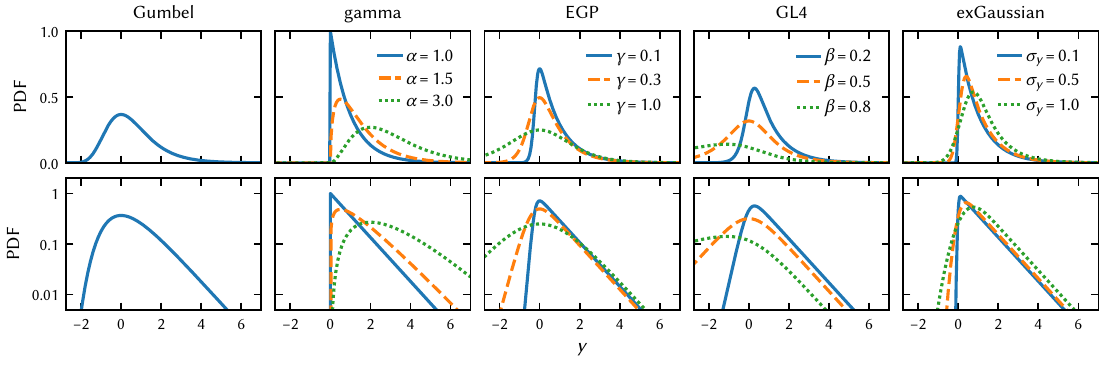}
    \caption{The probability distribution function (PDF) of each distribution discussed in sec.\,\ref{sec:dists}, plotted on both a linear (top) and log (bottom) scale. We hold $\mu$ and $\delta$ fixed across all examples, but present several choices of shape parameter in each case.}
    \label{fig:dists}
\end{figure*}

We find the Gumbel distribution to be an unappealing choice because its only parameters are its location ($\mu$) and scale ($\delta$): the probability distribution function (PDF) is given by
\begin{align}
    f_y(y) &= e^{- (y + e^{-y})} \, .
\end{align}
The shape of this function is presented in Fig.\,\ref{fig:dists}.

The limiting behavior of the magnitude PDF at bright and dim magnitudes is
\begin{align}
    \lim_{M \rightarrow -\infty} f_M(M) & \propto e^{M/\delta} 
    \label{eq:gumbelplus}
    \\
    \lim_{M \rightarrow +\infty} f_M(M) & \propto e^{-e^{M / \delta}} \, . \label{eq:gumbel12}
\end{align}
We obtain the correct behavior at bright magnitudes if 
\begin{align}
    \delta &= 1/\ln r \, .
\end{align}
However, we see from eq.\,\ref{eq:gumbel12} that the limiting behavior at dim magnitudes is also determined by $r$. There is no reason to expect that the behavior at dim magnitudes is governed solely by the population index, and so we conclude that a Gumbel distribution is an inappropriate choice for describing the meteor magnitude distribution.

\subsection{Gamma distribution}
\label{sec:gamma}

\cite{vida20} use a gamma distribution to characterize meteor magnitudes:
\begin{align}
    f_y(y) &= \frac{1}{\Gamma(\alpha)} 
    \begin{cases}
        y^{\alpha - 1} e^{-y} & M < \mu \\
        0 & M \ge \mu
    \end{cases} \, , \text{ where } \, \alpha > 0 \, .
\end{align}
The gamma distribution is the only distribution discussed in this section that is non-zero over a semi-infinite domain. That is, ${f_M(M) > 0}$ only when ${M < \mu}$.

When ${\alpha = 1}$, the gamma distribution is identical to an exponential distribution. Otherwise, the limiting behavior is
\begin{align}
    \lim_{M \rightarrow -\infty} f_M(M) & \propto e^{M/\delta} \\
    \lim_{M \rightarrow \mu} f_M(M) & \propto \left( 
        (\mu - M) / \delta
    \right)^{\alpha - 1} \, .
\end{align}
We see that a gamma distribution also resembles eq.\,\ref{eq:nm1} at bright magnitudes when ${\delta = 1/\ln r}$. It is also an improvement over a Gumbel distribution in that it has an additional shape parameter, $\alpha$, and so the behavior of the distribution at dim magnitudes is not completely determined by $r$. 

Note that \cite{vida20} do not equate $\delta$ and ${\ln r}$. Instead, they compute $r$ from the slope of the log of the distribution (specifically, the survival function) at a location that is one magnitude brighter than the inflection point. At this point, $y$ and $f_M$ are equal to
\begin{align}
    y_\text{ref} &= \alpha - 1 + \sqrt{\alpha - 1} + 1/\delta \\
    f_y'(y_\text{ref}) &= \frac{\delta^{-1} + \sqrt{\alpha-1}}{\delta^{-1} + \sqrt{\alpha-1} + \alpha-1}
    \label{eq:rvida}
\end{align}
If we set ${\alpha = 2}$ and ${\delta^{-1} = \ln 2.7}$, then eq.\,\ref{eq:rvida} is equal to 0.67. In other words, the apparent population index at the \citeauthor{vida20}\ reference point is approximately 2/3 as large as its asymptotic value.

\subsection{Exponentiated generalized Pareto (EGP) distribution}
\label{sec:egpd}

\cite{sotolongo08} and \cite{betzler15} propose that meteoroid masses follow what they call a nonextensive distribution \citep{tsallis88}, but instead appears to be a generalized Pareto distribution. The magnitude distribution then follows an exponentiated generalized Pareto (EGP) distribution \citep{lee18}, whose PDF can be written:
\begin{align}
    f_y(y) &= \frac{1}{\gamma} e^{y/\gamma} \left( 1 + \gamma^{-1} e^{y/\gamma} \right)^{-(1 + \gamma)} \,
        \text{,\, where }  \, \gamma > 0 \,.
\end{align}
The limiting behavior of this distribution resembles an exponential distribution at both ends of the magnitude distribution:
\begin{align}
    \lim_{M \rightarrow -\infty} f_M(M) & \propto e^{ M/\delta} \\
    \lim_{M \rightarrow +\infty} f_M(M) & \propto e^{- M/(\gamma \delta)} \, .
\end{align}
We again require ${\delta = 1/\ln r}$ in order to match the behavior of eq.\,\ref{eq:nm1} at bright magnitudes. The value of $\gamma$ can then be chosen to match the shape of the distribution at dim magnitudes.

In sec.\,\ref{sec:apply}, we will see that this distribution fits the data nearly as well as an exGaussian distribution. On the other hand, it can be unwieldy to work with: for instance, the mean of the distribution is a rather complicated combination of hypergeometric functions of $\gamma$.

We also find fault with the reasoning behind the distribution. \cite{sotolongo08} invoked a nonextensive distribution to describe the \emph{true} distribution of meteoroid masses prior to entering the atmosphere. This distribution was assumed to deviate from a power law at some size where particles tended to fragment due to collisions that occurred prior to the formation of comets. This argument neglects the fact that a turnover in the magnitude distribution is seen in all networks, regardless of what particle masses they are capable of detecting.

\subsection{Type IV generalized logistic (GL4) distribution}
\label{sec:gld4}

\cite{wu05} models the distribution of visual Leonid meteors by multiplying the PDF of an exponential distribution by a truncated hyperbolic tangent function \citep[their fit to the averaged perception probability data of][]{koschack90b}. \cite{trigo22} use their approach to create an idealized distribution of visual Leonid meteor magnitudes. \cite{moorhead24} also model the perception probability using a logistic function, which is equivalent to an (untruncated) hyperbolic tangent function.

The product of a logistic function and an exponential function is the PDF of a type IV generalized logistic (GL4) distribution whose shape parameters have a sum of unity:
\begin{align}
    f_y(y) &= \frac{1}{\beta \, B(1-\beta, \, \beta)} \frac{e^{-y}}{1+e^{-y/\beta}} \, \text{,\, where} \, 0 < \beta < 1 \, .
\end{align}
Here, $B$ is the beta function. Like the EGP distribution, this distribution resembles an exponential distribution at both extremes:
\begin{align}
    \lim_{M \rightarrow -\infty} f_M(M) & \propto e^{ M/\delta} \\
    \lim_{M \rightarrow +\infty} f_M(M) & \propto e^{-(M/\delta) (1-\beta)/\beta} \, .
\end{align}
We again require ${\delta = 1/\ln r}$ in order to match the limiting behavior at bright magnitudes to eq.\,\ref{eq:nm1}. The value of $\beta$ can then be used to match the behavior at dim magnitudes.

The properties of this distribution are discussed at greater length in \cite{moorhead24report}, who use it to model the solar longitude distribution of shower meteors.

\subsection{ExGaussian distribution}

An exGaussian distribution is the result of convolving a normal distribution with an exponential distribution \citep{grushka72}, and has the following PDF:
\begin{align}
    f_y(y) &= \frac{1}{2} \,
        e^{\sigma_y^2/2 - y} \,
        \text{erfc} \left[ \frac{\sigma_y^2 - y}{\sqrt{2} \, \sigma_y} \right] \, 
        \text{,\, where} \, \sigma_y > 0.
        \label{eq:fyexg} 
\end{align}
Here, the function ``erfc'' is the complementary error function.

The limiting behavior is that of an exponential distribution at bright magnitudes and a normal distribution at dim magnitudes:
\begin{align}
    \lim_{M \rightarrow -\infty} f_M(M) & \propto e^{M/\delta} \\
    \lim_{M \rightarrow +\infty} f_M(M) & \propto e^{-(M-\mu)^2 / 2 \sigma^2} \, ,
\end{align}
where ${\sigma = \delta \sigma_y}$.
As with all distributions discussed in this section, the exGaussian distribution requires ${\delta = 1/\ln r}$ to match the behavior of eq.\,\ref{eq:nm1} at bright magnitudes.

In eq.\,\ref{eq:fyexg}, we have written the PDF of an exGaussian in an unusual format; we did this in order to preserve as many parallels as possible with the other distributions in this section. The PDF is more typically written
\begin{align}
    f_x(x) = 
        \frac{\tau}{2} 
        &\exp \left( -\tau (x - \mu) + \frac{\tau^2 \sigma^2}{2} \right) \times \nonumber \\
        &\text{erfc} \left( 
            \frac{-\tau(x-\mu) + \tau^2 \sigma^2}{\sqrt{2} \, \tau \sigma}
        \right) \, .
\end{align}
We will use the expression 
\begin{align}
    x &\sim \text{ExG} \left( \tau, \mu, \sigma \right)
\end{align}
to indicate that the variable $x$ follows an exGaussian distribution with rate parameter $\tau$ (which is equivalent to $1/\delta$), location parameter $\mu$, and shape parameter ${\sigma = \sigma_y/\tau}$.

\section{Derivation}
\label{sec:deriv}

In this section, we describe how an exGaussian magnitude distribution can arise naturally from measurement error and/or variations in detection threshold. We begin by writing eq.\,\ref{eq:nm1} in the form of a probability distribution function (PDF):
\begin{align}
    f_M(M) &= \ln r \begin{cases}
        e^{(M-M_0) \ln r} & M < M_0 \\
        0 & M \ge M_0
    \end{cases} \, \label{eq:fM}
\end{align}
where $M_0$ is an upper limit on meteor magnitude whose value is unimportant to this derivation. This represents the true distribution of meteor magnitudes.

Digital cameras measure brightness by counting photons, and therefore the measured magnitude will differ from the true magnitude due to shot or Poisson noise. Additional differences may result from effects such as atmospheric scintillation or variations in the meteors' spectra. Let's use $m$ to denote the measured magnitude. Meteors will be detected only if they are brighter than the instrument's detection threshold: that is, if ${m < m_t}$.

Let's assume for the moment that the detection threshold, $m_t$, does not vary and that its value is known. So long as ${m_t \ll M_0 - \varepsilon}$, we find that
\begin{align}
    f_{m}(m) &= \ln r \begin{cases}
        e^{(m - m_t) \ln r} & m < m_t \\
        0 & m \ge m_t
    \end{cases} \, , 
    \label{eq:fMbar}
\end{align}
where $\varepsilon$ is the standard deviation of the magnitude error. Equation~\ref{eq:fMbar} applies when measurement error is the result of shot noise and/or normally distributed magnitude errors that occur prior to detection (see appendix~\ref{apx:a}). In other words, random errors in magnitude do not affect the observed magnitude distribution, so long as they occur prior to applying the detection threshold.

Now suppose that additional errors in magnitude occur \emph{after} detection; that is, these errors do not influence whether or not the meteor is detected. For example, perhaps the detection threshold applies to the brightest pixel, but the magnitude estimate is obtained by integrating the meteor's signal over the point spread function (PSF), which spans multiple pixels. We would then expect additional random error in the total magnitude beyond that which occurs in the brightest pixel. 
Let's assume that the additional error follows a normal distribution:
\begin{align}
    f_{m' \lvert m}(m') &= 
    \frac{1}{\sqrt{2 \pi} \epsilon} 
    e^{-(m' - m)^2/2 \epsilon^2}
\end{align}
If we multiply the above equation by eq.\,\ref{eq:fMbar} and integrate over $m$, we obtain the following PDF for $m'$:
\begin{align}
    f_{m'}(m') &= \frac{\rho}{2}
    e^{\rho (m' - m_t) + \frac{\rho^2 \epsilon^2}{2}} \,
    \text{erfc} \left(\frac{m' - m_t + \rho \epsilon^2}{\sqrt{2} \epsilon}
    \right) \, .
    \label{eq:fg1}
\end{align}
In other words,
\begin{align}
    -m' &\sim \text{ExG} \left( 
        \rho, 
        -m_t,
        \epsilon
    \right) \, .
\end{align}

On the other hand, suppose that there is no error introduced after detection, but that the detection threshold itself is variable:
\begin{align}
    f_{m_t}(m_t) &= 
    \frac{1}{\sqrt{2 \pi} \sigma} 
    e^{-(m_t - m_T)^2/2 \sigma^2}
\end{align}
If we multiply this equation by eq.\,\ref{eq:fMbar} and integrate over $m_t$, we obtain
\begin{align}
    f_{m}(m) &\propto 
    \frac{\rho}{2}
    e^{\rho (m - m_T)} \,
    \text{erfc} \left(\frac{m - m_T}{\sqrt{2} \sigma}
    \right)
\end{align}
which is again the PDF of an exGaussian distribution, albeit one that is shifted to the left by ${\rho \varepsilon^2}$ compared to eq.\,\ref{eq:fg1}. In other words,
\begin{align}
    -m &\sim \text{ExG} ( 
        \rho, \,
        -m_T - \rho \sigma^2, \,
        \sigma
    ) \, .
\end{align}
This shift occurs because more meteors are detected when the threshold is more sensitive: a variation in the detection limit therefore results in a magnitude distribution that is shifted to dimmer magnitudes compared to that produced by random errors in magnitude that are introduced after or independently of detection.

Finally, let's assume that the threshold is variable \emph{and} that post-detection measurement errors are present. In this case,
\begin{align}
    -m' &\sim \text{ExG} \left( 
        \rho, \,
        -m_T - \rho \sigma^2, \,
        \sqrt{\epsilon^2 + \sigma^2}
    \right) \, .
\end{align}

We would like to emphasize that this derivation assumes that the variation in detection threshold is normally distributed. This assumption is most likely to hold for instrumental meteor magnitude: that is, the brightness of a meteor as seen by the camera. Thus, the use of instrumental magnitudes should eliminate or at least minimize systematic variations in detection threshold.

\section{Application}
\label{sec:apply}

In this section we test our ability to fit an exGaussian data to two data sets: [1] faint optical meteors detected by EMCCD cameras \citep{gural22} at the Canadian Automated Meteor Observatory (CAMO), and [2] radar meteor echoes detected by the Canadian Meteor Orbit Radar \citep[CMOR;][]{webster04}. In each case, we sample or adjust the data to a common detection threshold.

\subsection{Faint optical meteors}
\label{sec:emccd}

Two pairs of electron-multiplying CCD (EMCCD) cameras with overlapping fields of view were added to CAMO's suite of instruments in 2016 \citep{brown20,gural22}. These cameras can detect meteors with magnitudes as dim as +8, making them almost as sensitive to small meteors as CMOR. The camera pipeline reports not only magnitude but also trajectory information; this allows us to exclude meteor showers, as their population indices can differ from that of the sporadic complex.

\subsubsection{Data cleaning}
\label{sec:clean1}

We have the best chance of constraining the population index when both the variation in detection threshold and the post-detection measurement error are small. We cannot control the latter, but we can take measures to reduce variation in detection threshold. The most important such measure is the use of apparent magnitude, not absolute (i.e., range-corrected) magnitude, for our analysis.

As mentioned above, we have selected only sporadic meteors for this analysis. Although multiple sources (helion/antihelion, apex, and north toroidal) are visible to CAMO, we have opted not to separate them. We made this choice for two reasons: [1] the sporadic sources have very similar population indices \citep{blaauw11} and [2] a larger data set is more likely to reveal deviations from a model.

We also limit our analysis to meteors seen at locations and times where and when the detection limit is similar. For instance, we select meteors whose peak is seen by a single camera at one of the stations (camera ``01G'') between January 2021 and December 2024, inclusive. We also noticed that the sensitivity is worse during twilight hours, including astronomical twilight. We therefore exclude meteors that peak in brightness near the edge of the sensor, and those that are seen when the solar elevation angle is above $-18^\circ$. (A detailed analysis of the detection threshold is the subject of the next paper in this series.) Finally, we require the location of the meteor at its peak brightness to be at least five frames away from the edge of the field of view of at least one camera, where we determine this using the apparent motion of the meteor per frame. This makes it more likely that we are truly capturing the meteor's peak magnitude. 
After these cuts, we are left with just over 20\,000 meteors: see table~\ref{tab:cuts}.

\begin{table}
    \centering
    \begin{tabular}{ll} \hline \hline
        & count \\ \hline
        Meteors seen in camera 01G & 32\,990 \\
        Solar elevation less than $-18^\circ$ & 29\,161 \\
        More than 5 frames' distance from sensor edge & 22\,913 \\
        Sporadic meteors & 20\,582
    \end{tabular}
    \caption{The number of meteors seen by camera 01G after each cut.}
    \label{tab:cuts}
\end{table}

\begin{figure}
    \centering
    \includegraphics[width=\linewidth]{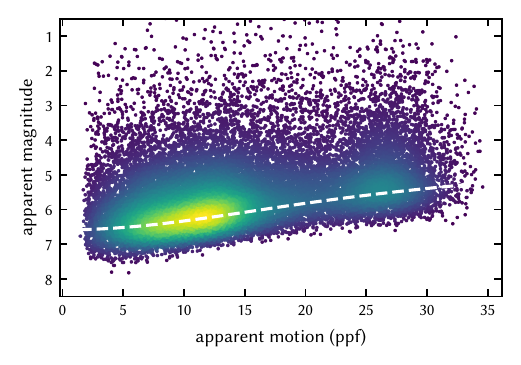}
    \caption{Peak apparent magnitude vs.\ apparent angular motion (in pixels per frame) for meteors detected by camera 01G. Points are color-coded by point density to help the reader identify the most populated areas (in yellow). The dashed white line follows eq.\,\ref{eq:dmag} with ${u_0 = 10}$~ppf.}
    \label{fig:dmag}
\end{figure}

The value of the meteor's brightest pixel is also affected by its angular motion relative to the camera station. The light produced by fast-moving meteors are smeared across a larger number of pixels, which can make it harder to detect. Rather than further restrict our data set, we have opted to correct for this effect using the following expression from \cite{molau16}:
\begin{align}
    \Delta m &= -2.5 \log_{10} \left[ 
        \frac{u_0}{u} \, \text{erfc} \left( 
            \frac{\sqrt{\pi}}{2} \frac{u}{u_0}
        \right)
    \right] \, . \label{eq:dmag}
\end{align}
This expression gives the predicted increase in the magnitude of the brightest pixel ($\Delta m$), where $u$ is the apparent motion of the meteor in pixels per frame and $u_0$ is a scaling factor. We find that ${u_0 = 10~\text{ppf}}$ fits the data well (see Fig.\,\ref{fig:dmag}). We subtract ${\Delta m}$ from the apparent magnitude to create what we term a ``motion-adjusted'' magnitude:
\begin{align}
    m_\text{adj} &= m - \Delta m \, .
\end{align}
This adjusted magnitude is effectively the instrumental magnitude, and is the quantity whose distribution we will model.

\subsubsection{Distribution fitting}
\label{sec:fit1}

We initially tried two different methods of distribution fitting: maximum likelihood estimation (MLE) and least-squares (we found the method of moments to be prohibitively difficult to implement for an EGP distribution). MLE and least-squares produced similar results for an exGaussian distribution, but least-squares performed far better for the gamma and Gumbel distributions. For this reason, we will focus on the least-squares results.

Our fitting process was as follows. First, we created a density histogram of motion-adjusted magnitude using Freedman–Diaconis binning \citep{freedman81}. The bin midpoints and densities served as our predictor and response values, respectively. Next, we minimized the square of the difference between the square root of the theoretical PDF at each bin midpoint and the square root of the corresponding response value. Taking the square root regularizes the variance and allows us to weight each data point equally \citep{johnson05,ohara10}. Technically, we should average the theoretical PDF across the bin \citep[see, e.g., sec.\,5.2 of][]{moorhead24report}, but we have enough data here that the difference in fit was insignificant. Sample code is presented in appendix~\ref{apx:code}.

\begin{figure}
    \centering
    \includegraphics[width=\linewidth]{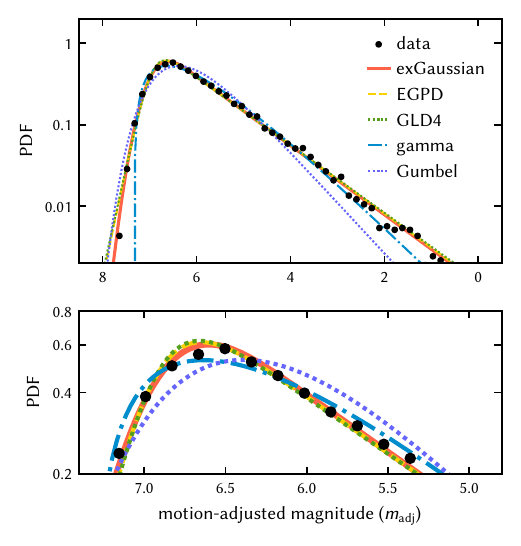}
    \caption{The PDF in each family that best fits the distribution of motion-adjusted magnitudes from one EMCCD camera (black points). The lower panel zooms in on the peak of the distribution.}
    \label{fig:fitmag}
\end{figure}

The best fit parameters for each distribution are given in table~\ref{tab:par1}, along with their estimated uncertainties and the reduced chi-squared value for each fit. We have converted the scaling parameter ($\delta$) to population index as follows:
\begin{align}
    r &= \exp (1/\delta) \\
    \sigma_r &= (r/\delta^2) \, \sigma_\delta
\end{align}
The best-fitting distributions are also shown in Fig.\,\ref{fig:fitmag}. An exGaussian distribution provides the closest fit to the data; this is easiest to see from the reduced chi-squared values. The EGP and GL4 distributions also fit the data very closely, although we can see that they overpredict the number of meteors at very dim magnitudes. The gamma and Gumbel distributions noticeably deviate from the data, and their fits does not meaningfully constrain the population index ($r$).

\begin{table}
    \centering
    \small
    \begin{tabular}{lcccc} \hline\hline
        & $r$ & $-\mu$ & shape & $\chi^2_\text{red}$ \\ \hline
        exGaussian & $2.7 \pm 0.3$ & $6.95 \pm 0.11$ & $\sigma = 0.28 \pm 0.07$ & 2.0 \\
        EGP       & $2.6 \pm 0.4$ & $6.64 \pm 0.16$ & $\gamma = 0.15 \pm 0.07$ & 2.9 \\
        GL4       & $2.6 \pm 0.4$ & $6.91 \pm 0.13$ & $\beta = 0.13 \pm 0.05$ & 3.3 \\
        gamma      & $4 \pm 2$ & $7.4 \pm 0.2$ & $\alpha = 2.1 \pm 0.7$ & 8.2 \\
        Gumbel     & $4.2 \pm 0.9$ & $6.36 \pm 0.12$ & -- &
        9.8 \\ 
    \end{tabular}
    \caption{Best-fit distribution parameters for EMCCD motion-adjusted magnitudes.}
    \label{tab:par1}
\end{table}

\subsection{Radar meteors}
\label{sec:cmor}

We also examine the magnitude (or rather, echo amplitude) distribution of single-station meteor echoes recorded by CMOR in the years 2016--2021 (inclusive). Because these data are single-station, we do not have radiants for these meteors and cannot identify which meteors are sporadic and which belong to showers. Instead, we minimize shower contamination by selecting meteors that are likely to be antihelion meteors, and excluding periods when major showers are active near the antihelion radiant. Focusing on one sporadic source makes it simpler to exclude periods of known shower contamination.

\subsubsection{Data cleaning}
\label{sec:clean}

We begin by selecting echoes that lie within $10^\circ$ of the antihelion echo line, where the antihelion radiant is assumed to be centered on Sun-centered ecliptic coordinates ${\lambda - \lambda_\odot = 202^\circ}$, ${\beta = 2.2^\circ}$ \citep{campbellbrown08}. We also require that the Sun is below the horizon at the time of the meteor's observation; this both excludes helion meteors and minimizes the effects of Faraday rotation.

\begin{figure}
    \centering
    \includegraphics[]{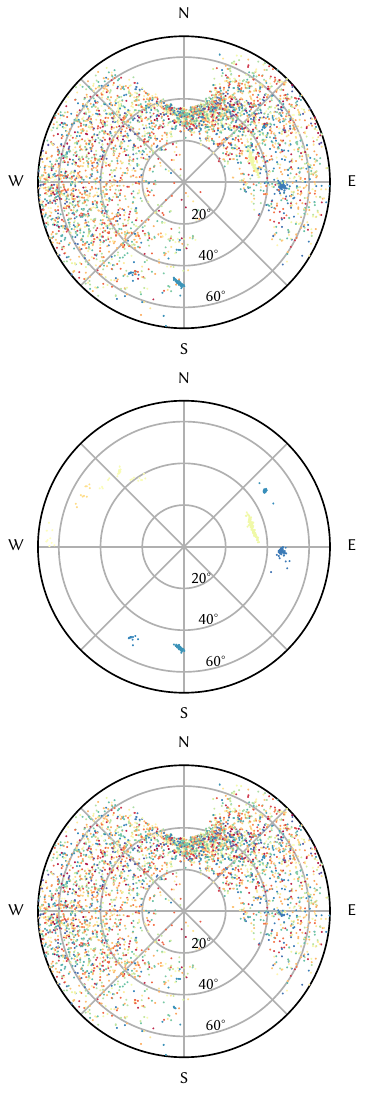}
    \caption{The azimuth and elevation angle of echoes recorded by CMOR for solar longitudes between $116^\circ$ and $118^\circ$. Points are color-coded by solar longitude within this range. Radial gridlines correspond to $20^\circ$ increments in zenith angle. In the top panel, we show all data. At center, we plot only cluster members identified by DBSCAN. At bottom, we plot the data that remains after we remove the clusters.}
    \label{fig:dbscan}
\end{figure}

The echo data are contaminated byfalse positives caused by events such as lightning. We excluded echoes with apparent elevation angles within $20^\circ$ of the horizon, as these data are more likely to be noise events. We then excluded additional noise events by identifying anomalously tight clusters in meteor azimuth, elevation, and time using DBSCAN \citep{ester96,schubert17}. We scanned the data in increments of two degrees in solar longitude and found that using azimuth and elevation angle ranks, rather than their values, naturally accounted for variations in background density. In Fig.\,\ref{fig:dbscan}, we show the event clusters that DBSCAN identified for a particularly ``noisy'' period from ${\lambda_\odot = 116^\circ}$ to $118^\circ$.

\begin{figure}
    \centering
    \includegraphics[width=\linewidth]{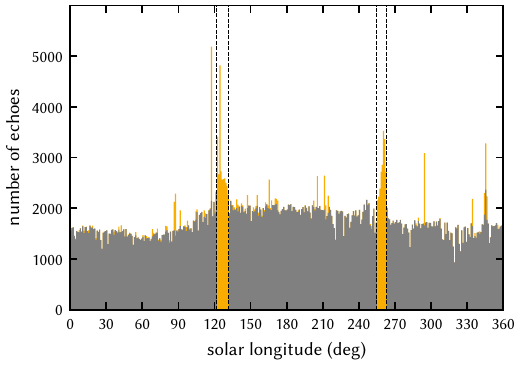}
    \caption{The distribution of solar longitudes in our data set. All data appear in yellow; anomalous events and two major meteor showers have been excluded from the data set shown in gray.}
    \label{fig:shist}
\end{figure}

The removal of noise clusters results in a smoother distribution of solar longitudes (see Fig.\,\ref{fig:shist}). However, two periods of time have noticeably elevated activity levels: these periods correspond to the Southern delta Aquariid (SDA, ${\lambda_\odot \approx 122^\circ-132^\circ}$) and Geminid (GEM, ${\lambda_\odot \approx 255^\circ-263^\circ}$) meteor showers. We exclude these time periods in order to focus on sporadic antihelion meteors. No other showers produce significant enhancements in the number of echoes, and thus shower contamination in the remaining data should be very slight.

\subsubsection{Radar amplitude}
\label{sec:amp}

A meteor's electron line density is typically assumed to be proportional to its brightness \citep{verniani73}. We do not directly measure the line density, however, but rather the echo amplitude. For an underdense radar meteor echo, the peak amplitude is approximately equal to
\begin{align}
    A &\simeq
        \frac{q}{q_\text{tr}}
        \frac{\sqrt{G_T \cdot G_R(\theta, \phi) \, P_T \, \zeta_R}}{4 \pi^2}
        \left( \frac{\lambda}{R} \right)^{3/2} \eta(R, n_a, v) \, ,
        \label{eq:amp}
\end{align}
where $q$ is the electron line density, $q_\text{tr}$ is a transitional electron line density, ${G_T \cdot G_R}$ is the total gain, ${P_T = \text{15 kW}}$ is the transmitter power, ${\zeta_R}$ is the receiver input impedance, ${\lambda = \text{10 m}}$ is the radar wavelength, and $R$ is the distance from the station to the meteor \citep[see eqs.\,8-6 and 8-7 of][]{mckinley61}. The factor $\eta$ is the attenuation due to observing biases \citep{ceplecha98} and depends on the initial trail radius ($n_a$) and meteor speed ($v$) as well as distance (since we have restricted our data set to include only nighttime observations, we ignore Faraday rotation). 
From eq.\,\ref{eq:amp}, we see that the amplitude is directly proportional to the line density, and so can function as an instrumental apparent magnitude.

In deriving eq.\,\ref{eq:amp}, we have made use of eq.\,8-30 of \cite{mckinley61}:
\begin{align}
    q_\text{tr} &= \frac{1}{\pi r_e}
    \label{eq:qtr}
\end{align}
where $r_e$ is the classical electron radius. Competing definitions of $q_\text{tr}$ exist, however: 
see eq.\,8-29 of \cite{mckinley61} and page~401 of \cite{ceplecha98} for examples.

The amplitude distribution is complicated by the fact that there is also an upper limit on the electron line density of underdense meteors. When the trail is sufficiently dense, the radar beam cannot penetrate into the trail and reflects only off its front surface. Such trails are called \emph{overdense} and the relationship between their density and the recorded amplitude (${A \propto q^{1/4}}$) differs from eq.\,\ref{eq:amp}. According to eq.\,8-32 of \cite{mckinley61}, meteors produce underdense-type echoes when
\begin{align}
    q &< q_\text{ov} = \pi^3 \left( \frac{r_0}{\lambda} \right)^2 q_\text{tr}
\end{align}
where $r_0$ is the initial trail radius. The transition from the underdense to the overdense regime may be gradual, and we therefore require ${q \ll q_\text{ov}}$.
In terms of amplitude, this is equivalent to
\begin{align}
    A &\ll A_\text{ov} \simeq
        \frac{\pi}{4} \sqrt{\frac{G_T \cdot G_R(\theta, \phi) \, P_T \, \zeta_R}{\lambda \, R^3}} \, r_0^2 
    \label{eq:aov}  
\end{align}
when we neglect the attenuation factor $\eta$.
We would like $A_\text{ov}$ to be as large as possible so that we can maximize the range of amplitudes for which the amplitude is larger than the detection threshold but smaller than ${A_\text{ov}}$.

\subsubsection{Distribution fitting}
\label{sec:fit2}

The radar gain pattern is a known function and the range can be measured, but unfortunately the initial trail radius is not well known. \cite{mckinley61} proposes that ${r_0 \propto n_a^{-1}}$, where $n_a$ is the atmospheric number density, and claims a ``slight velocity dependence.'' In contrast, \cite{ceplecha98} claim that ${r_0 \propto n_a^{-0.25} v^{0.6}}$ is a better fit to experimental data. \cite{jones05} found a similar dependence on number density, but a reversed relationship with speed: ${r_0 \propto n_a^{-0.2826} v^{-0.2}}$. Given this uncertainty in $r_0$, the dependence of the attenuation factor on $r_0$, the existence of several competing definitions for $q_\text{tr}$, and the gradual transition from underdense to overdense behavior, we will not attempt to determine $A_\text{ov}$ directly. 

\begin{figure}
    \centering
    \includegraphics{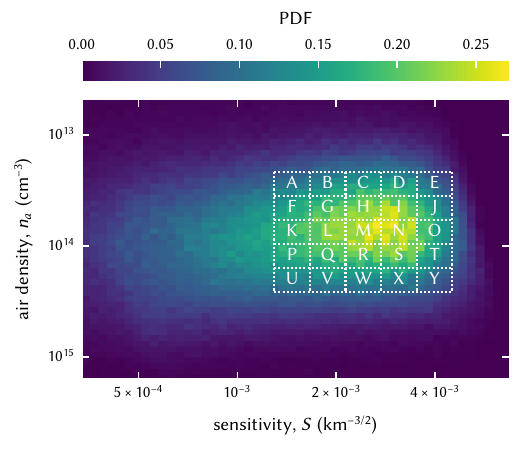}
    \caption{Two-dimensional density plot of our sensitivity factor (eq.\,\ref{eq:sens}) and atmospheric density, both on logarithmic scales, for probable antihelion meteors observed by CMOR. Notice that we have inverted the vertical axis so that low densities/high altitudes lie at the top of the plot. The grid overlay shows how we bin the data for analysis; each bin is labeled with a unique letter.}
    \label{fig:grid1}
\end{figure}

Instead, we separate $A_\text{ov}$ into two factors, $n_a$ and $S$, where $S$ is a ``sensitivity factor'' that incorporates $A_\text{ov}$'s dependence on view angle and range:
\begin{align}
    A_\text{ov} &\propto S \, n_a^{-2 \xi} \, ~ \text{, where} \\
    S &= \sqrt{\frac{G_T \cdot G_R(\theta, \phi)}{R^3}} \, ~ 
    \text{and} \label{eq:sens} \\
    r_0 &= n_a^{-\xi} \, .
\end{align} 
We have dropped all constants, including constants that are specific to CMOR. We bin the data by our two factors: Fig.\,\ref{fig:grid1} superimposes our grid on top of a heat map of the data. We size the bins so that their vertical span is
\begin{align}
    \Delta \ln n_a &= \frac{1}{2} \Delta \ln S \, ;
\end{align}
this aspect ratio corresponds to the expected dependence of $A_\text{ov}$ on $S$ and $n_a$ if ${\xi = 0.25}$ \citep[or ${r_0 \propto n_a^{-0.25}}$;][]{ceplecha98}.
To generate this plot, we calculated the atmospheric number density for each meteor according to the NRLMSISE00 model \citep{picone02} using the nrlmsise00 Python package.\footnote{\url{https://pypi.org/project/nrlmsise00/}} 

\begin{figure*}
    \centering
    \includegraphics{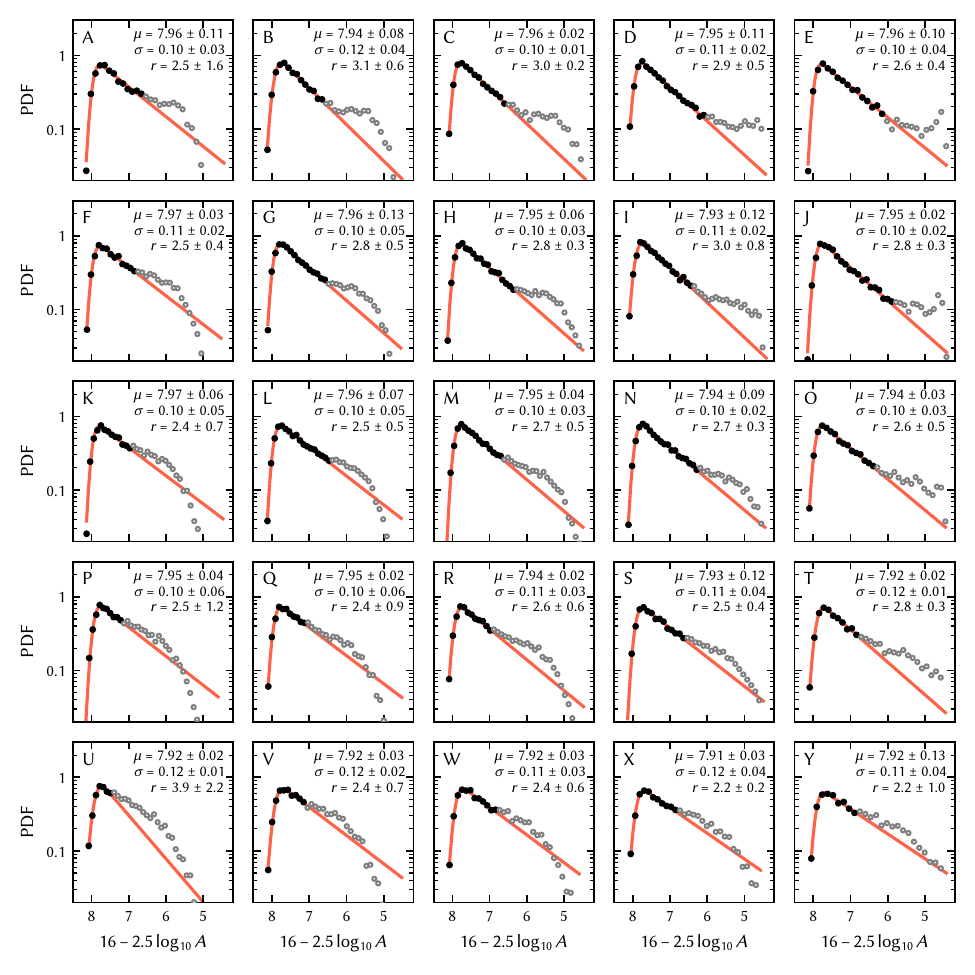}
    \caption{The distribution of antihelion meteor radar amplitudes detected by CMOR within each bin shown in Fig.\,\ref{fig:grid1}. The labels match those shown in Fig.\,\ref{fig:grid1} and we have arranged the plots so that they appear in the same order, with the lowest atmospheric densities (highest altitudes) in the top row and the large values of the sensitivity factor ($S$) in the right column.
    We have fit an exGaussian distribution to each subset. The solid black points are included in the fits, while the gray circles were excluded. Each plot is annotated with the parameters of the best-fitting exGaussian distribution.}
    \label{fig:grid2}
\end{figure*}

From eq.\,\ref{eq:aov}, we expect the largest separation between the detection threshold and $A_\text{ov}$ when $S$ is large and $n_a$ is small. Figure\,\ref{fig:grid2} seems to bear this out; at low atmospheric densities (top row), the data appear to follow an exGaussian distribution with an additional ``hump'' that presumably corresponds to overdense and transitional echoes. The location of this hump shifts to the right as $S$ increases (from left to right).

\begin{figure}
    \centering
    \includegraphics[width=\linewidth]{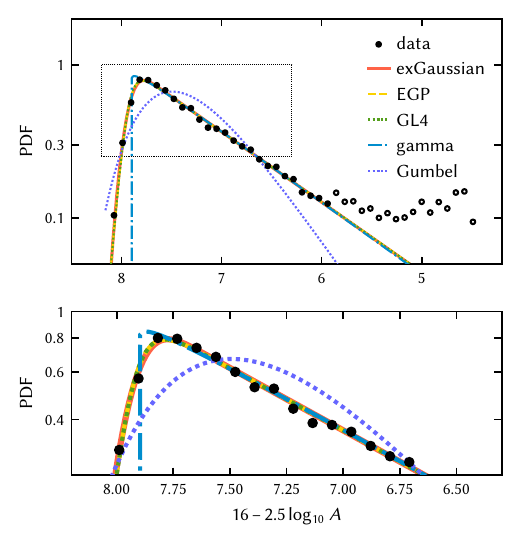}
    \caption{The distribution of antihelion meteor radar amplitudes detected by CMOR with low atmospheric densities and high sensitivity factors (sample Z). We have opted to display the PDF rather than the more commonly shown survival function. We struggled to fit the data with either a Gumbel or gamma distribution, but an exGaussian distribution fit the data very well at low amplitudes (as did the EGP and GL4 distributions). Open circles indicate that the data point in question has been excluded from our least-squares fitting.}
    \label{fig:fitcmor}
\end{figure}

Note that we have transformed the amplitude as follows:
\begin{align}
    a &= 16 - 2.5 \log_{10} A \, .
\end{align}
This quantity has the same expected scaling (${\delta = 1/ \ln r}$) as magnitude, and has been shifted so that the turnover in its distribution is similar to CMOR's limiting magnitude of +8 \citep{moorhead24}. This allows us to visually compare the range of magnitudes represented in figs.\,\ref{fig:fitmag} and \ref{fig:fitcmor}. However, $a$ does not convert amplitude to magnitude per meteor, as the relationship between the two quantities depends on the meteor's speed, initial radius, polarization angle, and the assumed ionization and luminous efficiencies \citep[see, e.g.,][]{weryk13}. 

The data in the top right bins should have the greatest separation between the detection limit and the underdense-overdense transition, and this is supported by Fig.\,\ref{fig:grid2}. We also notice that the overdense ``hump'' is sharper or more pronounced at low atmospheric densities. The reason for this is not entirely clear, but could arise due to the initial trail radius effect. At low atmospheric densities, the initial radius of the trail can be comparable to or larger than the radar wavelength, resulting in destructive interference between the radiation reflected from the near and far trail surfaces. This shifts the amplitude of these meteors to lower values (leftward) without affecting the amplitude of overdense meteors, creating a more distinct separation between the two populations. 

Because the amplitude distribution deviates from an exGaussian at large amplitudes, we do not use the entire dataset for our fits. We determined the cutoff value for our fits by looking for the point at which the deviation from an exGaussian distribution and the fit uncertainty in $r$ are both low. At low atmospheric densities/high altitudes, we were able to use a large fraction of the amplitude range. At low altitudes, on the other hand, we were able to use very little of the data, leading to large fit uncertainties. Furthermore, the overlap between underdense and overdense meteors at low altitudes/high atmospheric densities results in shallower estimates of the population density ($r$); in other words, when these populations are poorly separated in the data, we expect measurements of $r$ to be erroneously low. The estimated detection threshold ($\mu$) and shape parameter ($\sigma$) are fairly constant across all selected ranges in altitude and sensitivity, however, as can be seen in Fig.\,\ref{fig:grid1}.

\begin{figure}
    \centering
    \includegraphics{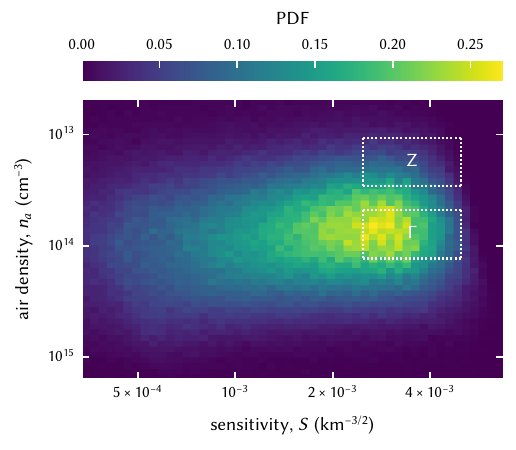}
    \caption{Two-dimensional density plot of our sensitivity factor (eq.\,\ref{eq:sens}) and atmospheric density, both on logarithmic scales, for probable antihelion meteors observed by CMOR. Two subsamples, ``Z'' and ``$\Gamma$,'' are circumscribed by the dotted lines and labeled.}
    \label{fig:gridzg}
    \vspace{2\baselineskip}

    \includegraphics{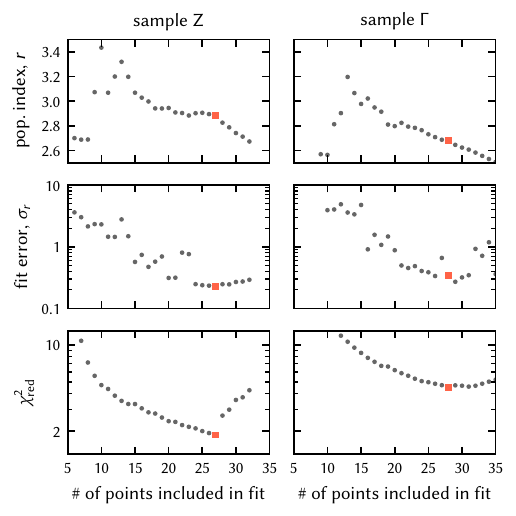}
    \caption{Fit diagnostics for samples Z and $\Gamma$. At left, we select meteors with large values of $S$ and low atmospheric densities (sample Z). We vary the number of points included in the fit and select the number that minimizes the fit uncertainty in population index ($\sigma_r$) and reduced chi-squared statistic ($\chi^2_\text{red}$); this selection is marked with a red square. At right, we select meteors with large values of $S$ and intermediate atmospheric densities (sample $\Gamma$). In this latter case, the minimum reduced chi-square and fit uncertainty in $r$ are both larger, and there is no region over which the best-fitting value of $r$ is relatively constant.}
    \label{fig:points}
\end{figure}

For our final fit, we select a larger sample from the high-sensitivity, low-atmospheric-density quadrant (see sample~Z in Fig.\,\ref{fig:points}). This sample contains 24,035 meteors, which is similar in size to our EMCCD sample (see table~\ref{tab:cut2}). We create an amplitude histogram for the data using the Freedman–Diaconis rule to automatically set the bin width. We find that the data begin to deviate from an exGaussian distribution in the 28th bin; when we fit an exGaussian distribution to the leftmost 27 histogram bins, we obtain the lowest fit error in $r$ and the lowest reduced chi-squared statistic (see left column of Fig.\,\ref{fig:points}). We are also encouraged to see that the best-fitting value of $r$ is a relatively stable function of the number of points included, $p$, for ${18 \lesssim p \lesssim 27}$. When too few points are included, the fit is of course less stable; when too many are included, the best-fitting value of $r$ decreases with $p$. In contrast, when we sample meteors at intermediate number densities, there is no range of $p$ over which the best-fitting value of $r$ is stable, hinting that overdense and underdense meteors cannot be cleanly separated (see right column of Fig.\,\ref{fig:points}).

\begin{table}
    \centering \small
    \begin{tabular}{lc} \hline \hline
        & count \\ \hline
        Within $10^\circ$ of antihelion echo line & 
            993\,370 \\
        When sun is below horizon &
            652\,251 \\
        After noise removal & 632\,695 \\
        Excluding SDA and GEM showers & 585\,708 \\
        Sample Z & \hphantom{0}24\,035
    \end{tabular}
    \caption{The number of radar meteors after each cut.}
    \label{tab:cut2}
\end{table}

Our best fits to sample~Z are summarized in table~\ref{tab:par2} and shown in Fig.\,\ref{fig:fitcmor}. We report the chi-squared statistic for each fit, but, since we have chosen the number of points to minimize $\chi^2_\text{red}$ for an exGaussian fit, this is not an entirely fair way to compare fits. Nevertheless, the relative values of $\chi^2_\text{red}$ confirm what we can tell by eye: that the exGaussian, EGP, and GL4 distributions are capable of fitting the data reasonably well, while the gamma and Gumbel distributions cannot. The exGaussian fit provides the most precise measurement of $r$, however. This value is a little higher than that obtained from the EMCCD data, but ${2.9 \pm 0.2}$ does not disagree with ${2.7 \pm 0.3}$.

\begin{table}
    \centering
    \small
    \begin{tabular}{lcccc} \hline\hline
        & $r$ & $\mu$ & shape & $\chi_\text{red}^2$ \\ \hline
        exGaussian & $2.9 \pm 0.2$ & $7.95 \pm 0.01$ & 
            $\sigma = 0.10 \pm 0.09$ & 1.9 \\
        EGP        & $2.9 \pm 1.1$ & $7.8 \pm 0.2$ & 
            $\gamma = 0.06 \pm 0.04$ & 3.2 \\
        GL4        & $2.9 \pm 0.8$ & $7.94 \pm 0.08$ & 
            $\beta = 0.06 \pm 0.05$ & 3.3 \\
        gamma      & $3 \pm 11000$ & $7.90 \pm 0.71$ & 
            $\alpha = 1 \pm 9000$ & 157 \\
        Gumbel     & $9 \pm 6$ & $7.5 \pm 0.4$ & -- & 86 \\
    \end{tabular}
    \caption{Best-fit distribution parameters and reduced chi-squared values ($\chi^2_\text{red}$) for underdense CMOR radar echoes from probable antihelion meteors.}
    \label{tab:par2}
\end{table}

\section{Conclusions}

If meteor magnitudes (or amplitudes) are subject to normally distributed variations in detection threshold or post-detection measurement errors, the resulting distribution of instrumental magnitudes is an exponentially modified Gaussian (exGaussian) distribution. We examined the magnitude distribution of faint optical meteors seen by a CAMO EMCCD camera and meteor echo amplitudes recorded by CMOR, and found that an exGaussian distribution provides a better fit to these data than previously proposed distributions.

Although the use of an exGaussian distribution to model meteor magnitudes is justified on the basis of its derivation and match to several data sets, we note that the exGaussian distribution also happens to be convenient to work with. For instance, it has been fully implemented as a continuous distribution in the SciPy Python package. Furthermore, its three parameters can be easily converted to physically meaningful quantities: population index, mean detection threshold, and a combination of detection threshold variation and measurement error.
For this reason, we recommend an exGaussian distribution over an exponentiated generalized Pareto (EGP) distribution and a type IV generalized logistic (GL4) distribution: although these latter two distributions fit the data almost as well, they are less interpretable and more difficult to work with.

The population index can therefore be estimated by fitting an exGaussian distribution to a set of observed meteor magnitudes. For our two data sets, we obtained estimates of ${r = 2.7 \pm 0.3}$ and ${r = 2.9 \pm 0.2}$. These translate to mass indices of ${s \simeq 2.0}$ to ${s \simeq 2.2}$, depending on the assumed relationship between magnitude and mass. Our estimates are consistent with recent literature, which yields values of $s$ ranging from 2.0 to 2.17 \citep{blaauw11,pokorny16,janches19,vida20,kipreos25}. 
While it is encouraging that we obtain population index estimates that are in keeping with the literature, the main purpose of this paper is to establish the exGaussian distribution as an appropriate fit for the observed distribution of meteor magnitudes. We will make use of this information in our next paper, in which we examine the meteor detection limit of CAMO's EMCCD cameras.

Our analysis of radar data shows that the measured population index is sensitive to contamination by overdense meteors, but, in contrast, estimates of the detection threshold and shape parameter are robust to this contamination. This suggests that exGaussian distribution fits provide reliable estimates of the detection threshold that can be used for applications such as flux calculations.

\bibliography{refs}

@article{barghini24,
    author = {{Barghini}, D. and {Battisti}, M. and
        {Belov}, A. and {Bertaina}, M. and
        {Bertone}, S. and {Bisconti}, F. and
        {Blaksley}, C. and {Blin}, S. and
        {Bolmgren}, K. and {Cambi{\`e}}, G. and
        {Capel}, F. and {Casolino}, M. and
        {Cellino}, A. and {Churilo}, I. and
        {Coretti}, A.~G. and {Crisconio}, M. and
        {De La Taille}, C. and {Ebisuzaki}, T.
        and {Eser}, J. and {Fenu}, F. and
        {Filippatos}, G. and {Franceschi}, M.~A.
        and {Fuglesang}, C. and {Gardiol}, D.
        and {Golzio}, A. and {Gorodetzky}, P.
        and {Kajino}, F. and {Kasuga}, H. and
        {Klimov}, P. and {Kungel}, V. and
        {Kuznetsov}, V. and {Manfrin}, M. and
        {Marcelli}, L. and {Mascetti}, G. and
        {Marsza{\l}}, W. and {Mignone}, M. and
        {Miyamoto}, H. and {Murashov}, A. and
        {Napolitano}, T. and {Ohmori}, H. and
        {Olinto}, A. and {Parizot}, E. and
        {Picozza}, P. and {Piotrowski}, L.~W.
        and {Plebaniak}, Z. and
        {Pr{\'e}v{\^o}t}, G. and {Reali}, E. and
        {Reynaud}, F. and {Ricci}, M. and
        {Romoli}, G. and {Sakaki}, N. and
        {Sharakin}, S. and {Shinozaki}, K. and
        {Szabelski}, J. and {Takizawa}, Y. and
        {Vagelli}, V. and {Valentini}, G. and
        {Vrabel}, M. and {Wiencke}, L. and
        {Zotov}, M.},
    title = "{Observation of meteors from space with
        the Mini-EUSO detector on board the
        International Space Station}",
    journal = {Astronomy and Astrophysics},
    year = 2024,
    volume = {687},
    eid = {A304},
    pages = {A304},
    doi = {10.1051/0004-6361/202449236},
}

@article{betzler15,
    author = {{Betzler}, A.~S. and {Borges}, E.~P.},
    title = "{Non-extensive statistical analysis of
        meteor showers and lunar flashes}",
    journal = {MNRAS},
    year = 2015,
    volume = {447},
    number = {1},
    pages = {765-771},
    doi = {10.1093/mnras/stu2426},
}

@article{blaauw11,
    author = {{Blaauw}, R.~C. and {Campbell-Brown},
        M.~D. and {Weryk}, R.~J.},
    title = "{Mass distribution indices of sporadic
        meteors using radar data}",
    journal = {MNRAS},
    year = 2011,
    volume = {412},
    number = {3},
    pages = {2033-2039},
    doi = {10.1111/j.1365-2966.2010.18038.x},
}

@article{blaauw16,
    author = {{Blaauw}, R.~C. and {Campbell-Brown},
        M. and {Kingery}, A.},
    title = "{Optical meteor fluxes and application
        to the 2015 Perseids}",
    journal = {MNRAS},
    year = 2016,
    volume = {463},
    number = {1},
    pages = {441-448},
    doi = {10.1093/mnras/stw1979},
}

@article{blaauw17,
    author = {{Blaauw}, R.~C.},
    title = "{The mass index and mass of the Geminid
        meteoroid stream as determined with
        radar, optical and lunar impact data}",
    journal = {P\&SS},
    year = 2017,
    volume = {143},
    pages = {83-88},
    doi = {10.1016/j.pss.2017.04.007},
}

@article{brown20,
    author = {{Brown}, Peter and {Weryk}, Robert J.},
    title = "{Coordinated optical and radar
        measurements of low velocity meteors}",
    journal = {Icarus},
    year = 2020,
    volume = {352},
    eid = {113975},
    pages = {113975},
    doi = {10.1016/j.icarus.2020.113975},
}

@article{browne56,
    author = {{Browne}, I.~C. and {Bullough}, K. and
        {Evans}, S. and {Kaiser}, T.~R.},
    title = "{Characteristics of radio echoes from meteor trails. II. The distribution of meteor magnitudes and masses}",
    journal = {Proceedings of the Physical Society B},
    year = 1956,
    volume = {69},
    number = {1},
    pages = {83-97},
    doi = {10.1088/0370-1301/69/1/311},
}

@article{campbellbrown08,
    author = {{Campbell-Brown}, M.~D.},
    title = "{High resolution radiant distribution
        and orbits of sporadic radar
        meteoroids}",
    journal = {Icarus},
    year = 2008,
    volume = {196},
    number = {1},
    pages = {144-163},
    doi = {10.1016/j.icarus.2008.02.022},
}

@article{campbellbrown16,
    author = {{Campbell-Brown}, M.~D. and {Blaauw},
        R. and {Kingery}, A.},
    title = "{Optical fluxes and meteor properties of
        the Camelopardalid meteor shower}",
    journal = {Icarus},
    year = 2016,
    volume = {277},
    pages = {141-153},
    doi = {10.1016/j.icarus.2016.05.001},
}

@article{ceplecha98,
    author = {{Ceplecha}, Zden{\v{e}}k and
        {Borovi{\v{c}}ka}, Ji{\v{r}}{\'\i} and
        {Elford}, W. Graham and {Revelle},
        Douglas O. and {Hawkes}, Robert L. and
        {Porub{\v{c}}an}, Vladim{\'\i}r and
        {{\v{S}}imek}, Milo{\v{s}}},
    title = "{Meteor phenomena and bodies}",
    journal = {Space Science Reviews},
    year = 1998,
    volume = {84},
    pages = {327-471},
    doi = {10.1023/A:1005069928850},
}

@article{egal20,
    author = {{Egal}, A. and {Wiegert}, P. and
        {Brown}, P.~G. and {Campbell-Brown}, M.
        and {Vida}, D.},
    title = "{Modeling the past and future activity
        of the Halleyid meteor showers}",
    journal = {A\&A},
    year = 2020,
    volume = {642},
    eid = {A120},
    pages = {A120},
    doi = {10.1051/0004-6361/202038953},
}

@article{ehlert20,
    author = {{Ehlert}, Steven and {Erskine},
        Rhiannon Blaauw},
    title = "{Measuring fluxes of meteor showers with
        the NASA All-Sky Fireball Network}",
    journal = {P\&SS},
    year = 2020,
    volume = {188},
    eid = {104938},
    pages = {104938},
    doi = {10.1016/j.pss.2020.104938},
}

@inproceedings{ester96,
    author = {{Ester}, Martin and {Kriegel},
        Hans-Peter and {Sander}, J{\"o}rg and
        {Xu}, Xiaowei},
    title = "{A density-based algorithm for
        discovering clusters in large spatial
        databases with noise}",
    booktitle = {Second International Conference on
        Knowledge Discovery and Data Mining},
    year = 1996,
    editor = {{Pfitzner}, D.~W. and {Salmon}, J.~K.},
    pages = {226-331},
}

@article{freedman81,
    author = {{Freedman}, David and {Diaconis}, Persi},
    title = "{On the histogram as a density estimator:
             \ensuremath{L_2} theory}",
    year = {1981},
    journal = {Zeitschrift f\"{u}r Wahrscheinlichkeitstheorie und Verwandte Gebiete},
    volume = {57},
    number = {4},
    pages = {453–476},
    DOI = {10.1007/bf01025868},
}

@article{grushka72,
    author = {{Grushka}, Eli},
    title = "{Characterization of exponentially modified Gaussian peaks in chromatography}",
    journal = {Analytical Chemistry},
    year = 1972,
    volume = {44},
    number = {11},
    pages = {1733-1738},
    doi = {10.1021/ac60319a011},
}

@article{gural22,
    author = {{Gural}, P. and {Mills}, T. and
        {Mazur}, M. and {Brown}, P.},
    title = "{Development of a very faint meteor
        detection system based on an EMCCD
        sensor and matched filter processing}",
    journal = {Experimental Astronomy},
    year = 2022,
    volume = {53},
    number = {3},
    pages = {1085-1126},
    doi = {10.1007/s10686-021-09828-3},
}

@article{hawkins58,
    author = {{Hawkins}, Gerald S. and {Upton},
        Edward K.~L.},
    title = "{The influx rate of meteors in the
        Earth's atmosphere.}",
    journal = {Astrophysical Journal},
    year = 1958,
    volume = {128},
    pages = {727},
    doi = {10.1086/146585},
}

@article{jacchia67,
    author = {{Jacchia}, Luigi and {Verniani}, Franco
        and {Briggs}, Robert E.},
    title = "{An analysis of the atmospheric
        trajectories of 413 precisely reduced
        photographic meteors}",
    journal = {Smithsonian Contributions to
        Astrophysics},
    year = 1967,
    volume = {10},
    pages = {1-139},
}

@article{janches19,
    author = {{Janches}, D. and {Brunini}, C. and
        {Hormaechea}, J.~L.},
    title = "{
        A decade of sporadic meteoroid mass
        distribution indices in the southern
        hemisphere derived from
        SAAMER{\textquoteright}s meteor
        observations}",
    journal = {Astronomical Journal},
    year = 2019,
    volume = {157},
    number = {6},
    eid = {240},
    pages = {240},
    doi = {10.3847/1538-3881/ab1b0f},
}

@book{johnson05,
    author = {{Johnson}, Norman L. and {Kemp}, Adrienne W. and {Kotz}, Samuel},
    title = {Univariate Discrete Distributions},
    publisher = {Wiley},
    year = {2005},
    pages = {168},
    doi = {10.1002/0471715816.ch4},
}

@article{jones05,
    author = {{Jones}, J. and {Campbell-Brown}, M.},
    title = "{The initial train radius of sporadic
        meteors}",
    journal = {MNRAS},
    year = 2005,
    volume = {359},
    number = {3},
    pages = {1131-1136},
    doi = {10.1111/j.1365-2966.2005.08972.x},
}

@article{kaiser60,
    author = {{Kaiser}, T.~R.},
    title = "{The determination of the incident flux
        of radio-meteors}",
    journal = {MNRAS},
    year = 1960,
    volume = {121},
    pages = {284},
    doi = {10.1093/mnras/121.3.284},
}

@article{kipreos25,
    author = {{Kipreos}, Yung E. and {Moorhead},
        Althea V. and {Brown}, Peter G. and
        {Campbell-Brown}, Margaret and {Cooke},
        William J.},
    title = "{Improved measurement of radar meteor
        shower mass indices}",
    journal = {Icarus},
    year = 2025,
    volume = {441},
    eid = {116652},
    pages = {116652},
    doi = {10.1016/j.icarus.2025.116652},
}

@article{koschack90a,
    author = {{Koschack}, R. and {Rendtel}, J.},
    title = "{Determination of spatial number density
        and mass index from visual meteor
        observations (I)}",
    journal = {WGN, Journal of the IMO},
    year = 1990,
    volume = {18},
    number = {2},
    pages = {44-58},
}

@article{koschack90b,
    author = {{Koschack}, R. and {Rendtel}, J.},
    title = "{Determination of spatial number density
        and mass index from visual meteor
        observations (II)}",
    journal = {WGN, Journal of the IMO},
    year = 1990,
    volume = {18},
    number = {4},
    pages = {119-140},
}

@article{koten23,
    author = {{Koten}, P. and {Shrben{\'y}}, L. and
        {Spurn{\'y}}, P. and {Borovi{\v{c}}ka},
        J. and {{\v{S}}tork}, R. and {Henych},
        T. and {Voj{\'a}{\v{c}}ek}, V. and
        {M{\'a}nek}, Jan},
    title = "{{\ensuremath{\tau}} Herculid meteor
        shower in the night of 30/31 May 2022
        and the meteoroid properties}",
    journal = {Astronomy and Astrophysics},
    year = 2023,
    volume = {675},
    eid = {A70},
    pages = {A70},
    doi = {10.1051/0004-6361/202346537},
}

@article{lee18,
    author = {{Lee}, Seyoon and {Kim}, Joseph H. T.},
    title = {Exponentiated generalized Pareto distribution: Properties and applications towards extreme value theory},
    year = {2018},
    journal = {Communications in Statistics - Theory and Methods},
    volume = {48}, 
    number = {8},
    pages = {2014–2038},
    doi = {10.1080/03610926.2018.1441418},
}

@book{mckinley61,
    author = {{McKinley}, Donald William Robert},
    title = "{Meteor Science and Engineering}",
    year = 1961,
    publisher = {McGraw-Hill},
    address = {New York, N.Y.},
    pages = {189}
}

@inproceedings{molau14,
    author = {{Molau}, Sirko and {Barentsen}, Geert
        and {Crivello}, Stefano},
    title = "{Obtaining population indices from video
        observations of meteors}",
    booktitle = {Proceedings of the International
        Meteor Conference, Giron, France, 18-21
        September 2014},
    year = 2014,
    editor = {{Rault}, J. -L. and {Roggemans}, P.},
    pages = {74-80},
}

@article{molau16,
    author = {{Molau}, S. and {Crivello}, S. and
        {Goncalves}, R. and {Saraiva}, C. and
        {Stomeo}, E. and {Kac}, J.},
    title = "{Results of the IMO Video Meteor Network:
        March 2016, and discussion about the
        meteor limiting magnitude}",
    journal = {WGN, Journal of the IMO},
    year = 2016,
    volume = {44},
    number = {4},
    pages = {120-126},
}

@article{molau22,
    author = {{Molau}, Sirko},
    title = "{Flux density determination in MetRec
        and MeteorFlux}",
    journal = {WGN, Journal of the IMO},
    year = 2022,
    volume = {50},
    number = {5},
    pages = {126-133},
}

@article{moorhead19,
    author = {{Moorhead}, Althea V. and {Egal},
        Auriane and {Brown}, Peter G. and
        {Moser}, Danielle E. and {Cooke},
        William J.},
    title = "{Meteor shower forecasting in near-Earth
        space}",
    journal = {JSR},
    year = 2019,
    volume = {56},
    pages = {1531-1545},
    doi = {10.2514/1.A34416},
}

@article{moorhead24,
    author = {{Moorhead}, Althea V. and {Vida}, Denis
        and {Brown}, Peter G. and
        {Campbell-Brown}, Margaret D.},
    title = "{A reference meteor magnitude for
        intercomparable fluxes}",
    journal = {AJ},
    year = 2024,
    volume = {168},
    number = {1},
    eid = {16},
    pages = {16},
    doi = {10.3847/1538-3881/ad496e},
}

@techreport{moorhead24report,
    author = {{Moorhead}, Althea V. and 
        {Campbell-Brown}, Margaret D.
        and {Brown}, Peter G.},
    title = "{The activity profiles and peak flux 
        of radar meteor showers}",
    year = 2024,
    institution = "NASA",
    number = "OSMA/MEO/Report–13",
}

@article{moorhead25,
    author = {{Moorhead}, Althea V. and {Cooke},
        William J. and {Brown}, Peter G. and
        {Campbell-Brown}, Margaret D.},
    title = "{The threshold at which a meteor shower
        becomes hazardous to spacecraft}",
    journal = {Advances in Space Research},
    year = 2025,
    volume = {75},
    number = {1},
    pages = {1145-1162},
    doi = {10.1016/j.asr.2024.08.012},
}

@article{ohara10,
    author = {{O'Hara}, Robert B. and {Kotze}, D. Johan},
    title = "{Do not log‐transform count data}",
    year = {2010},
    journal = {Methods in Ecology and Evolution},
    volume = {1},
    number = {2},
    pages = {118–122},
    doi = {10.1111/j.2041-210x.2010.00021.x},
}

@article{picone02,
    author = {{Picone}, J.~M. and {Hedin}, A.~E. and
        {Drob}, D.~P. and {Aikin}, A.~C.},
    title = "{NRLMSISE-00 empirical model of the
        atmosphere: Statistical comparisons and
        scientific issues}",
    journal = {Journal of Geophysical Research (Space
        Physics)},
    year = 2002,
    volume = {107},
    number = {A12},
    eid = {1468},
    pages = {1468},
    doi = {10.1029/2002JA009430},
}

@article{pokorny16,
    author = {{Pokorn{\'y}}, P. and {Brown}, P.~G.},
    title = "{A reproducible method to determine the
        meteoroid mass index}",
    journal = {Astronomy and Astrophysics},
    year = 2016,
    volume = {592},
    eid = {A150},
    pages = {A150},
    doi = {10.1051/0004-6361/201628134},
}

@article{schubert17,
    author = {{Schubert}, Erich and {Sander}, J\"{o}rg and {Ester}, Martin and {Kriegel}, Hans-Peter and {Xu}, Xiaowei},
    title = "{DBSCAN revisited, revisited: Why and how you should (still) use DBSCAN}",
    journal = {ACM Transactions on Database Systems},
    year = 2017,
    volume = {42},
    no = {3},
    eid = {19},
    doi = {10.1145/3068335},
}

@article{shustov22,
    author = {{Shustov}, B.~M. and {Zolotarev},
        R.~V.},
    title = "{Mass indices of meteoric bodies. I. Formation model of meteoroid streams}",
    journal = {Astronomy Reports},
    year = 2022,
    volume = {66},
    number = {2},
    pages = {179-189},
    doi = {10.1134/S1063772922020093},
}

@article{sotolongo08,
    author = {{Sotolongo-Costa}, Oscar and 
               {Gamez}, R. and
               {Luz\'{o}n}, F. and
               {Posadas}, A. and
               {Beckmann}, Pablo Weigandt},
    title = "{Non extensivity in meteor showers}",
    journal = {Apeiron},
    year = 2008,
    volume = {15},
    number = {2},
    pages = {187-201},
    eid = {2-s2.0-47249159518}
}

@article{trigo22,
    author = {{Trigo-Rodr{\'\i}guez}, Josep M. and
        {Blum}, J{\"u}rgen},
    title = "{Learning about comets from the study of
        mass distributions and fluxes of
        meteoroid streams}",
    journal = {MNRAS},
    year = 2022,
    volume = {512},
    number = {2},
    pages = {2277-2289},
    doi = {10.1093/mnras/stab2827},
}

@article{tsallis88,
    author = {{Tsallis}, Constantino},
    title = {Possible generalization of Boltzmann-Gibbs statistics},
    year = {1988},
    journal = {Journal of Statistical Physics},
    volume = {52},
    number = {1–2},
    pages = {479–487},
    doi = {10.1007/bf01016429}
}

@article{verniani73,
    author = {{Verniani}, Franco},
    title = "{An analysis of the physical parameters
        of 5759 faint radio meteors}",
    journal = {JGR},
    year = 1973,
    volume = {78},
    number = {35},
    pages = {8429-8462},
    doi = {10.1029/JB078i035p08429},
}

@article{vida20,
    author = {{Vida}, D. and {Campbell-Brown}, M. and
        {Brown}, P.~G. and {Egal}, A. and
        {Mazur}, M.~J.},
    title = "{A new method for measuring the meteor
        mass index: Application to the 2018
        Draconid meteor shower outburst}",
    journal = {A\&A},
    year = 2020,
    volume = {635},
    eid = {A153},
    pages = {A153},
    doi = {10.1051/0004-6361/201937296},
}

@article{vida22,
    author = {{Vida}, Denis and {Blaauw Erskine},
        Rhiannon C. and {Brown}, Peter G. and
        {Kambulow}, Jonathon and
        {Campbell-Brown}, Margaret and {Mazur},
        Michael J.},
    title = "{Computing optical meteor flux using
        Global Meteor Network data}",
    journal = {MNRAS},
    year = 2022,
    volume = {515},
    number = {2},
    pages = {2322-2339},
    doi = {10.1093/mnras/stac1766},
}

@article{watson39,
    author = {{Watson}, Fletcher},
    title = "{Influences of limiting magnitude upon
        meteor frequency}",
    journal = {PNAS},
    year = 1939,
    volume = {25},
    number = {5},
    pages = {243-245},
    doi = {10.1073/pnas.25.5.243},
}

@article{webster04,
    author = {{Webster}, A.~R. and {Brown}, P.~G. and {Jones}, J. and {Ellis}, K.~J. and {Campbell-Brown}, M.},
    year = 2004,
    title = "{Canadian Meteor Orbit Radar (CMOR)}",
    journal = {Atmospheric Chemistry \& Physics},
    volume = {4},
    number = {3},
    pages = {679-684},
    doi = {10.5194/acpd-4-1181-2004},
}

@article{weryk13,
    author = {{Weryk}, Robert J. and {Brown}, Peter
        G.},
    title = "{Simultaneous radar and video
        meteors II: Photometry and
        ionisation}",
    journal = {Planetary \& Space Science},
    year = 2013,
    volume = {81},
    pages = {32-47},
    doi = {10.1016/j.pss.2013.03.012},
}

@article{wu05,
    author = {{Wu}, Guang-jie},
    title = "{An analytical function of the
        probabilities of perception and its
        application to the calculation of the
        population index in the visual
        observation of meteors}",
    journal = {New Astronomy},
    year = 2005,
    volume = {11},
    number = {1},
    pages = {68-79},
    doi = {10.1016/j.newast.2005.05.006},
}

\newpage
\appendix

\section{The influence of Poisson noise on the distribution of measured meteor magnitudes}
\label{apx:a}

Let's define ${y = -M \ln r}$. The distribution of $y$ is then:
\begin{align}
    f_y(y) &= \begin{cases}
        e^{-(y-y_0)} & y > y_0 \\
        0 & y \le y_0
    \end{cases}
\end{align}
Let $\nu$ be the expected number of photons produced by the meteor that are intercepted by the camera during one observation period. This quantity will be proportional to the meteor's brightness:
\begin{align}
    \nu &= \beta e^{y - y_0}
\end{align}
where $\beta$ is a constant of proportionality. The distribution of $\nu$ can be easily derived using the change-of-variables technique:
\begin{align}
    f_\nu(\nu) &= \begin{cases}
        \dfrac{\beta}{\nu^2} & \nu > \beta \\[8pt]
        0 & \nu \le \beta
    \end{cases}
\end{align}
The observed number of photons follows a Poisson process:
\begin{align}
    f_{n \lvert \nu}(n) &= \begin{cases}
        \dfrac{e^{-\nu} \nu^n}{n!} & n \ge 0 \\[8pt]
        0 & n < 0
    \end{cases}
\end{align}
The marginal distribution can be obtained as follows:
\begin{align}
    f_n(n) &= \int_\beta^\infty f_{n \lvert \nu}(n) \, f_\nu(\nu) \, d \nu \\
    &= \frac{\beta}{n(n-1)} \, Q(n-1, \, \beta)
\end{align}
where $Q$ is the regularized upper incomplete gamma function.
Finally, the observed brightness is given by
\begin{align}
    \bar{y} &= y_0 + \ln (n/\beta)
\end{align}
and therefore
\begin{align}
    f_{\bar{y}}(\bar{y}) &= \frac{\beta}{\beta e^{\bar{y}-y_0} - 1} \, Q(\beta e^{\bar{y}-y_0} - 1, \, \beta)
\end{align}
When the number of photons ($n = \beta e^{\bar{y}-y_0}$) is large, this distribution reduces to 
\begin{align}
    f_{\bar{y}}(\bar{y}) &\sim e^{-(\bar{y}-y_0)} \, .
    \label{eq:fby}
\end{align}

\newpage
\section{Python demonstration}
\label{apx:code}

\begin{figure}[h]
\includegraphics{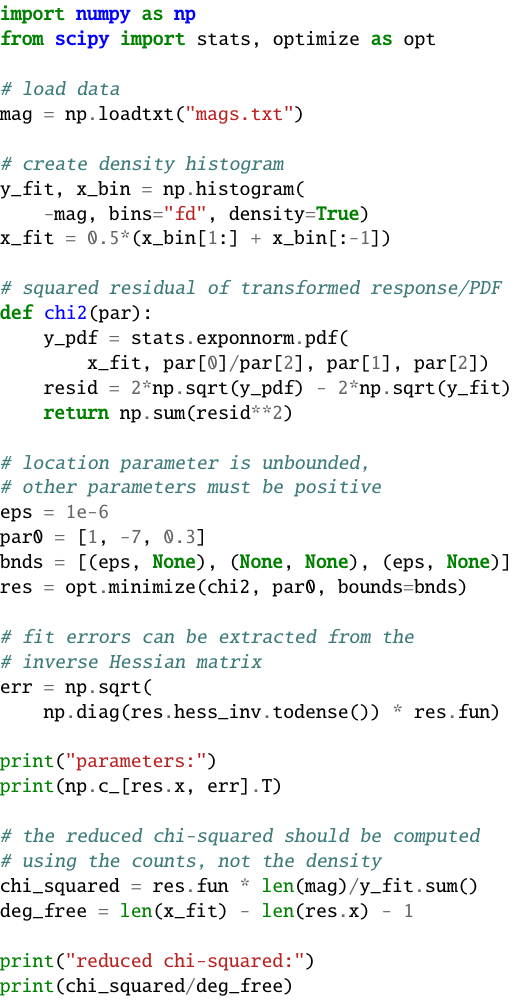}
\caption{Fitting an exGaussian distribution to magnitude data in Python.}
\label{fig:code}
\end{figure}

Figure\,\ref{fig:code} demonstrates how to fit an exGaussian distribution to magnitude data in Python using the least-squares approach. Notice that the exGaussian distribution is called ``exponnorm'' in the SciPy package, that the shape parameter ($k$) differs from ours ($\sigma_y$):
\begin{align}
    k &= \delta / \sigma \, ,
\end{align}
and that SciPy refers to $\sigma$ as the scale (not our $\delta$). We show how to convert between the two formalisms in the function ``chi2.''

\vfill

\vfill

\end{document}